\documentclass[11pt,a4paper]{article}

\usepackage[symbol]{footmisc}

\usepackage{jheppub}

\usepackage{slashed}
\usepackage{diagbox}
\usepackage{subfigure}
\usepackage{amsmath}
\usepackage{dutchcal}
\usepackage{mathrsfs}
\usepackage{amssymb}
\usepackage{bbm}

\usepackage{amsthm}
\usepackage{amsfonts}
\usepackage{dsfont}
\usepackage{amssymb, bm}
\usepackage{array}
\usepackage{booktabs}
\usepackage{overpic}
\usepackage{txfonts}

\def\II{\hbox{{1}\kern-.25em\hbox{l}}}

\newcommand{\beq}{\begin{equation}}
\newcommand{\eeq}{\end{equation}}
\newcommand{\bal}{\begin{align}}
\newcommand{\eal}{\end{align}}
\newcommand{\x}{{\bm x}}
\newcommand{\y}{{\bm y}}

\newcommand{\nn}{\nonumber}

\newcommand \widebar [1] {\overline{#1}}

\makeatletter
\newcommand{\newparallel}{\mathrel{\mathpalette\new@parallel\relax}}
\newcommand{\new@parallel}[2]{%
  \begingroup
  \sbox\z@{$#1T$}
  \resizebox{!}{\ht\z@}{\raisebox{\depth}{$\m@th#1/\mkern-5mu/$}}%
  \endgroup
}
\makeatother

\allowdisplaybreaks

\newcounter{MBQ}

\title{Total Gluon Helicity from Lattice without Effective Theory Matching}

\author[a,c]{Zhuoyi Pang,}

\author[b,a]{Fei Yao,}

\author[a,b]{and Jian-Hui Zhang}

\subheader{
\begin{flushright}
\end{flushright}
}

\affiliation[a]{School of Science and Engineering, The Chinese University of Hong Kong, Shenzhen 518172, China}

\affiliation[b]{
   Center of Advanced Quantum Studies, Department of Physics, Beijing Normal University, Beijing 100875, China}
\affiliation[c]{University of Science and Technology of China, Hefei, Anhui, 230026, P.R.China}

\emailAdd{pangzhuoyi@cuhk.edu.cn}
\emailAdd{feiyao@mail.bnu.edu.cn}
\emailAdd{zhangjianhui@cuhk.edu.cn}

\abstract{
We propose two approaches for extracting the total gluon helicity contribution to proton spin from lattice QCD, one from local operator matrix elements in a fixed gauge accessible on lattice with feasible renormalization, and the other from gauge-invariant nonlocal gluon correlators. Neither of these approaches requires a matching procedure when converted to the $\overline {\rm MS}$ scheme. Our proposal resolves a long-standing inconsistency in the literature regarding lattice calculations of the total gluon helicity, and has the potential to greatly facilitate these calculations.
}


\begin{document}

\maketitle


\section{Introduction}
\label{SEC:Introduction}
Since the European Muon Collaboration (EMC) experiment measured the quark spin contribution to proton spin through deep inelastic scattering with polarized muon and proton, the internal spin structure of the proton has been a subject of extensive study in QCD for many years. Now it is well-established that the quark spin contributes about $\sim 30\%$ of the proton spin~\cite{deFlorian:2009vb,Nocera:2014gqa,Ethier:2017zbq}, and the remaining contributions come from the gluon spin and the orbital angular momenta of quarks and gluons. Among these contributions, the understanding of the gluon spin has posed a theoretical challenge, mainly due to the absence of a local gauge invariant gluon spin operator. Nevertheless, it can be measured in polarized high-energy scattering experiments by probing the spin-dependent gluon helicity distribution
\beq
\Delta g(x)=g_+(x)-g_-(x),
\eeq
where $g_\pm(x)$ denotes the density of gluons in the longitudinally polarized proton with momentum fraction $x$ and the same/opposite helicity as the proton. The integral of $\Delta g(x)$ over the momentum fraction $x$, denoted as $\Delta G$ in the literature, gives the total gluon helicity contribution to the proton spin.

In field theory, the gluon helicity distribution is defined as the proton matrix element of a non-local gauge invariant operator~\cite{Manohar:1990kr}
\beq\label{eq:deltag}
\Delta g(x)=\frac{i}{2 x P^+}\int \frac{d\xi^-}{2\pi} e^{-ix\xi^- P^+}\left<PS\right|F_a^{+\mu}(\xi^-)\mathcal L_{ab}(\xi^-,0)\tilde F_{b,\mu}^+(0)\left|PS\right>,
\eeq
where $\tilde{F}^{\alpha \beta}=1/2\epsilon^{\alpha\beta\mu\nu}F_{\mu\nu}$ is the dual field strength, $\mathcal L$ is the lightcone gauge link, $a,b$ denote the color indices, and $\xi^\pm=(\xi^0\pm\xi^z)/\sqrt 2$ are lightcone coordinates. Gauge invariance dictates that it must be the field strength, not the gauge field that appears in the above expression, and the gauge link is inserted to compensate for the gauge transformations at two different spacetime points. 
After integrating over $x$, one obtains
\beq\label{eq:deltaG}
\Delta G=\int dx \frac{i}{2 x P^+}\int \frac{d\xi^-}{2\pi} e^{-ix\xi^- P^+}\left<PS\right|F_a^{+\mu}(\xi^-)\mathcal L_{ab}(\xi^-,0)\tilde F_{b,\mu}^+(0)\left|PS\right>.
\eeq
In the lightcone gauge $A^+=0$, the above expression becomes a local operator and has a clear physical interpretation, where it coincides with the gluon spin $(\vec E_a\times \vec A_a)^z$.
 
The question of how to understand the physical meaning of $\Delta G$ without going to a specific gauge like the lightcone gauge, has sparked a lot of studies in the past few decades~(for a recent review, see,  e.g., \cite{Liu:2021lke,Ji:2020ena}). In particular, in Ref.~\cite{Ji:2013fga} it has been shown that the clear physical interpretation of $\Delta G$ is intimately connected to the boost to the infinite momentum frame (IMF). Actually, $\Delta G$ can be obtained by boosting the matrix element of a gauge-invariant but frame-dependent gluon spin operator, $\vec E\times \vec A_\perp$, to the IMF, where $\vec A_\perp$ is the physical part of the gauge field that has the same transformation properties as the field strength under gauge transformations. Therefore, the gluon helicity can be understood as the IMF limit of the gauge invariant part of gluon spin. The physical part of the gauge field $\vec A_\perp$ takes a complicated nonlocal form in general, but it reduces to $\vec A$ in Coulomb gauge. This fact provides a practical possibility for nonperturbative calculations of the gluon helicity using lattice QCD. Although the nonlocal lightcone correlation in Eq.~(\ref{eq:deltaG}) cannot be directly accessed on lattice, we can calculate the matrix element of $\vec E\times \vec A_\perp$, or equivalently the matrix element of $\vec E\times \vec A$ in Coulomb gauge, in a proton state with a finite but large momentum. After an appropriate renormalization, the result can be converted to $\Delta G$ using a factorization or matching formula presented in Refs.~\cite{Ji:2013fga,Ji:2014lra}, which is the first example of the large-momentum effective theory (LaMET)~\cite{Ji:2013dva,Ji:2014gla,Ji:2020ect} factorization. If we denote the matrix element $\langle PS|\vec E \times \vec A_\perp|PS\rangle$, or equivalently $\langle PS|\vec E \times \vec A|PS\rangle_{C.G.}$ (the subscript $C. G.$ denotes the Coulomb gauge fixing) as $\Delta \tilde G$, then the factorization takes the following form~\cite{Ji:2014lra}
\beq\label{eq:deltaGfact}
\Delta \tilde G={\cal C}_{gg} \Delta G +{\cal C}_{gq} \Delta \Sigma + h.t.,
\eeq
where $\Delta\Sigma$ is the quark spin contribution, $h.t.$ denotes power contributions suppressed by the proton momentum, and the matching coefficients are given up to the next-to-leading (NLO) order as~\cite{Ji:2014lra}
\beq
{\cal C}_{gg}=1+{a_s C_A}\frac 7 3\ln \frac{P_z^2}{\mu^2}+fin.,\ \ \ \ {\cal C}_{gq}={a_s C_F}\frac 4 3\ln \frac{P_z^2}{\mu^2}+fin.,
\eeq
where $a_s=\alpha_s/(4\pi)$, the proton momentum is $P_\mu=(P_0,0,0,P_z)$, $\mu$ is the renormalization scale in the $\overline{\rm MS}$ scheme, and $fin.$ denotes finite pieces. Note that $\Delta\Sigma$ can be calculated from the matrix element $\langle PS| \bar\psi \gamma^\mu\gamma_5\psi|PS\rangle$, thus Eq.~(\ref{eq:deltaGfact}) allows us to extract $\Delta G$ from $\Delta\tilde G$. Later on, it was found~\cite{Hatta:2013gta} that the above calculation of using the gauge-fixed matrix element of $\vec E \times \vec A$ can be generalized from Coulomb gauge to a universality class of gauges that leaves the transverse components of the gauge field intact under a Lorentz boost.

Another possibility of calculating the total gluon helicity is to extract the gluon helicity distribution $\Delta g(x)$ using LaMET, and then take its first moment. In LaMET, one starts from an appropriately chosen equal-time Euclidean nonlocal gluon correlation, known as the quasi-light-front (quasi-LF) correlation. For the gluon helicity distribution, there are several possible choices for the quasi-LF correlation~\cite{Zhang:2018diq}. Here we take the following as an example
\beq
\tilde h(z, P_z)=\langle PS|F^{3\mu}(z){\cal L}(z,0){\tilde F}^0_\mu(0) |PS\rangle,
\eeq
where $z$ is the Euclidean distance along the direction of $P_z$.
After an appropriate renormalization and Fourier transform, it is turned to a momentum space quasi-distribution (the subscript $R$ denotes the renormalized quasi-LF correlation)
\beq
\Delta {\tilde g}(x)=\int\frac{dz}{2\pi x P_z}e^{i x z P_z}\tilde h_R(z, P_z),
\eeq
which can be mapped to the gluon helicity distribution through the following LaMET factorization~\cite{Ji:2020ect} 
\beq\label{eq:deltagfact}
\Delta \tilde g(x)=\int \frac{dy}{|y|} C_{gg}\Bigg(\frac{x}{y},\frac{\mu}{y P_z}\Bigg)\, \Delta g(y)+\int \frac{dy}{|y|} C_{gq}\Bigg(\frac{x}{y},\frac{\mu}{y P_z}\Bigg)\, \Delta q(y) + h.t.,
\eeq
where $\Delta q(y)$ is the quark helicity distribution, and the latest calculation of the matching coefficients in a state-of-the-art scheme up to the NLO has been presented in Refs.~\cite{Yao:2022vtp, Ma:2022gty}. For illustration purpose, here we only show the results in the physical region
\begin{align}
C_{gg}&=\delta\left(1-\frac{x}{y}\right)+4a_sC_A  \, \left\{ \frac{(2 x^2-3 x y+2 y^2)}{(x-y)\, y} \left(\ln \frac{\mu ^2}{4y^2P_z^2}-\ln\frac{ x (y-x)}{y^2}\right)  + fin. \right\}, ~~~~~ 0<x<y<1,\nn\\
C_{gq}&=\delta\left(1-\frac{x}{y}\right)+4a_sC_F  \, \left\{ \frac{(x-2y)}{y}\left(\ln \frac{\mu ^2}{4y^2P_z^2}-\ln\frac{x (y-x)}{y^2}\right)  + fin. \right\}, ~~~~~~~~~~~~~~~~~~~~0<x<y<1,
\end{align}
as the logarithmic momentum dependence $\ln(\mu/P_z)$ appears only in the physical region.

However, there is an inconsistency in the two approaches above. According to LaMET, the intrinsic momentum scale in the quasi-observables is the parton momentum $y P_z$ rather than the hadron momentum $P_z$. Such a momentum scale is then traded for the renormalization scale of the lightcone observables by the matching. This shall also be true in the factorization of the total gluon helicity Eq.~(\ref{eq:deltaGfact}). But in Eq.~(\ref{eq:deltaGfact}) only the total parton helicity is involved, and there is no momentum fraction at all. To put it differently, the multiplicative factorization form in Eq.~(\ref{eq:deltaGfact}) cannot be reproduced in general from Eq.~(\ref{eq:deltagfact}) by integrating over the momentum fraction $x$ on both sides. What is typically generated is a convoluted factorization form, unless the matching becomes trivial where the intrinsic parton momentum scale dependence drops out. Moreover, all existing proposals on calculating $\Delta G$ from the matrix element of $\vec E\times \vec A$ in a fixed gauge suffer from practical difficulties, either the nonperturbative renormalization is difficult to carry out or the desired gauge cannot be achieved on the lattice. For example, in Ref.~\cite{Yang:2016plb} the calculation was done in Coulomb gauge, where a nonperturbative renormalization was difficult to implement. The authors chose to match the bare result to the continuum through lattice perturbation theory, the convergence of which is known to be rather poor.

To resolve these issues, here we propose two different approaches. One is based on local operator matrix elements in a fixed gauge accessible on lattice with feasible renormalization, and the other is based on gauge invariant nonlocal gluon quasi-LF correlations. It turns out that by choosing appropriate local or nonlocal operators, the matching can be made trivial and the two factorization formulas above become consistent, provided that the relevant quantities are converted to the $\overline{\rm MS}$ scheme. Our proposal has the potential to greatly facilitate lattice calculations of the total gluon helicity.

The rest of the paper is organized as follows: in Sec.~\ref{background}, we discuss the gauge dependence of the topological current $K^{\mu}$, and the subtlety in taking its forward matrix element. Based on a boost argument, we propose to calculate its nonforward matrix element in Coulomb gauge on lattice, and explain how to do the renormalization in the regularization-independent momentum subtraction scheme (RI/MOM)~\citep{Martinelli:1994ty}. We then show how the total gluon helicity can be extracted from this matrix element without a perturbative matching. 
In Sec.~\ref{SEC:matchnonforward}, we present our second approach based on gauge-invariant nonlocal gluon correlators, and show how they can be connected to the total gluon helicity, again without a perturbative matching. 
We then conclude in Sec.~\ref{SEC:conclusion}.

\section{Gluon helicity from local operator matrix elements in a fixed gauge} \label{background}
As discussed in the Introduction, the existing proposals of calculating the total gluon helicity from the proton matrix element of $\vec{E}\times\vec{A}$ in a specific gauge suffers from the difficulty of implementing either the gauge or the nonperturbative renormalization. In this section, we give a new proposal relating the total gluon helicity to the local topological current in a fixed gauge, which circumvents the above problems and does not require a perturbative matching in the $\overline{\text{MS}}$ scheme.
 
\subsection{A brief introduction of topological current}
The EMC measurement of the spin asymmetry in polarized muon-proton scattering in the 1980s indicated that the total quark spin content was comparable to zero, in contradiction with the naive quark model expectations~\citep{EuropeanMuon:1987isl}. It was then proposed that due to the Adler-Bell-Jackiw anomaly~\citep{Adler:1969gk,Bell:1969ts}, besides quark helicity, the forward matrix element of the axial vector current will also include the gluon helicity content, providing a possible way out of the ``proton spin crisis" ~\citep{Altarelli:1988nr,Carlitz:1988ab,Jaffe:1989jz,Efremov:1989cb,Efremov:1989sn}. Moreover, in the parton model where the lightcone gauge is a natural choice, the total gluon helicity has a direct relation to the topological current $K^+$~\citep{Manohar:1990kr,Jaffe:1995an}:
\begin{equation}
    \big<PS\big|K^{+}\big|PS\big>=4S^+\Delta G,
\end{equation}
where $S^\mu=(P_z,0,0,P_0)$ is the spin vector of the proton, and in the calculation of $\Delta G$ from Eq.~\eqref{eq:deltaG} the principal-value prescription of the $1/x$ pole and a boundary condition $A^{i}(\infty^-)=0$ have been assumed. The first attempt of calculating $\big<PS\big|K^{\mu}\big|PS\big>$ on lattice was done in the temporal gauge\citep{Mandula:1990ce}, where the periodic boundary condition was adopted. However, the constructed $K^{\mu}$ on lattice shared few important properties with $K^{\mu}$ in the continuum\citep{Liu:1995kb}.

The topological current is defined as:
\begin{align} \label{eq:topo}
K^{\mu}=&\epsilon^{\mu\nu\rho\sigma}\text{Tr}\big[A_{\nu}F_{\alpha\beta}-\frac{2}{3}ig_{s}A_{\nu}A_{\alpha}A_{\beta}\big] =\frac{1}{2}\epsilon^{\mu\nu\rho\sigma}\left(A_{\nu}^{a}F_{\rho\sigma}^{a}+\frac{g_s}{3}f_{abc}A_{\nu}^{a}A_{\rho}^{b}A_{\sigma}^{c}\right),
\end{align}
where after the first equal sign we take trace of the product of color matrices. The four-divergence of $K^{\mu}(x)$ gives the topological charge density $Q(x)$: $\frac{\alpha_s}{4\pi}\partial_{\mu}K^{\mu}(x)=Q(x)$, with $Q(x)=\frac{\alpha_s}{8\pi}F_{\mu\nu}^{a}\tilde{F}^{\mu\nu,a}(x)$. 
The integration of $Q(x)$ over space-time gives the winding number of QCD vacuum: $\int d^4xQ(x)=\nu~ (\nu\in \text{integer})$. Under the gauge transformation, we have:
\begin{align} \label{eq:gauge_K}
A_{\mu}\rightarrow & UA_{\mu}U^{\dagger}-\frac{i}{g_s}U\partial_{\mu}U^{\dagger},\nonumber \\
    K_{\mu}\rightarrow &
    K_{\mu}-\frac{2i}{g_s}\varepsilon^{\mu\nu\alpha\beta}\partial_{\alpha}\text{Tr}\left[(\partial_{\nu}U^{\dagger})UA_{\beta}\right]+\frac{2}{3g_s^2}\varepsilon^{\mu\nu\alpha\beta}\text{Tr}\left[U^{\dagger}(\partial_{\nu}U)U^{\dagger}(\partial_{\alpha}U)U^{\dagger}(\partial_{\beta}U)\right] \nonumber \\
=&K^{\mu}+\delta K_1^{\mu}+\delta K_2^{\mu}.
\end{align}
The second term in the last line of Eq.~\eqref{eq:gauge_K}, $\delta K_1^{\mu}$, is a total derivative and vanishes in the forward matrix element. $\delta K_2^{\mu}$ comes from the gauge transformation of the three-gluon term in Eq.~\eqref{eq:topo}, which is sensitive to the topology of gluon field configurations. For example, at the boundary of 4D Euclidean space, it can be written as a Jacobian of the mapping from hypersphere $\mathbb{S}^3$ to the Non-Abelian gauge group~\citep{Belavin:1975fg,Crewther:1978zz}. Only for local gauge transformations, $\delta K_2^{\mu}$ can be written as a total derivative with zero divergence~\citep{Cronstrom:1983af}, whose forward matrix element vanishes.  Thus for a gauge and indice choice with nonvanishing three-gluon term in $K^{\mu}$, both forward and non-forward matrix elements of $K^{\mu}$ will be gauge-dependent. 

In addition, the forward matrix element of $K^\mu$ entails some subtlety that can be best elucidated by examining its non-forward matrix element, whose general decomposition reads:
\begin{equation} \label{eq:general}
    \big<P'S\big|K^{\mu}\big|PS\big>=V_{i}^{\mu}a_{i}+\tilde{V}_{j}^{\mu}b_{j},
\end{equation}
where the sum over $i~(j)$ is implicit, $a_{i}$ and $b_{i}$ are, in general, gauge-dependent functions. $V_i^{\mu}$~($\tilde{V}_i^{\mu}$) can be any four-vector appearing in the matrix element or the gauge condition. For example, in covariant gauges, under the requirement of parity, together with constraints from time reversal and hermiticity, we have:
\begin{equation} \label{eq:covariant}
    \big<P'S\big|K^{\mu}\big|PS\big>_{\partial\cdot A=0}=S^{\mu} a_{1}(q^{2})+q^{\mu} b_{1}(q^{2}),
\end{equation}
with $q^{\mu}=P^{\mu}-P'^{\mu}$ being the momentum transfer.

As shown in~\citep{Manohar:1990eu,Balitsky:1991te}, the $b_i$ in Eq.~(\ref{eq:general}) contains a gauge-dependent massless pole, 
making the forward limit dependent on the direction along which the momentum transfer approaches zero. In covariant gauges, we have $b_1(q^2)\sim{1}/{q^2}$. In the massive Schwinger model and covariant gauge, Kogut and Susskind have explicitly shown~\citep{Kogut:1974kt} that the unphysical pole comes from the coupling between $K^{\mu}$ and a ghost field. The ghost field, together with a scalar field with equal but opposite metric, forms a Goldstone dipole. They will cancel out with each other in physical matrix elements~\citep{Kogut:1974kt}. The discussion in QCD is more involved, but the massless ghost pole still exists in the propagator $\big<0\big|K^{\mu}K^{\nu}\big|0\big>$, and thus in $b_i$~\citep{Veneziano:1979ec}. 

Some authors identify $\text{lim}_{q^2\rightarrow0}\,a_1(q^2)$ as the total gluon helicity~\citep{Efremov:1989cb,Efremov:1989sn,Shore:1999be}, which is doubtful to us for two reasons. Firstly, unlike an axial gauge, in the IMF limit the covariant gauge does not approach the lightcone gauge~\citep{Hatta:2013gta}. In the axial gauge, $a_{1}(0)$ differs from $\Delta G$ by a power-suppressed contribution in the large momentum frame~\citep{Balitsky:1991te}, while in covariant gauges $a_1(0)$ cannot be related to the total gluon helicity due to their different boost behavior. Second, the three-gluon term in $K^\mu$ does not vanish in covariant gauges. Thus in covariant gauges, $a_1(0)$ in general differs from $\Delta G$ by an unknown nonperturbative quantity~\citep{Bass:1997zz,Bass:2004xa}.

\subsection{The nonforward matrix element of $K^{\mu}$ in the Coulomb gauge}
There are two sets of gauges that are most often used in lattice calculations: the Landau gauge and the Coulomb gauge. From the above discussion, we have seen that in covariant gauges, there is a distinction between the matrix element of $K^{\mu}$ and the total gluon helicity. In this subsection, we explore the possibility of relating $\big<K^{\mu}\big>$ to the total gluon helicity, in the Coulomb gauge ($\big<\cdots\big>$ denotes the nonforward matrix element of   ``$\cdots$'' in the subsequent discussion unless stated otherwise).
We justify that in the large momentum frame, the matrix element of $K^{\mu}$ in the Coulomb gauge differs from that in the lightcone gauge by a higher-twist contribution.  Moreover, an efficient renormalization procedure is proposed, by utilizing the perturbative gauge-invariance of the forward limit of $\big<K^{\mu}\big>$.

In Ref.~\cite{Hatta:2013gta}, it has been shown that the total gluon helicity Eq.~\eqref{eq:deltaG} can be written as the following matrix element
\begin{equation} \label{eq:Aphys}
    \Delta G=\frac{1}{2P^+}\big<PS\big|\big(\vec{E}\times\vec{A}_{\text{phys}}\big)^z\big|PS\big>,
\end{equation}
with $\vec{E}^{i}=F^{i+}, A_{\text{phys}}^{\mu}={F^{+\mu}}/{D^+}$. 
In the lightcone gauge, $A_{\text{phys}}^{\mu}$ reduces to $A^{\mu}$.
In contrast, at finite momentum different 
proposals have been made on defining the physical part (or the transverse part) of the gluon field (denoted as $A_{\perp}^{\mu}$ hereafter)~\citep{Chen:2008ag,Ji:2013fga}. For example, in Ref.~\citep{Ji:2013fga}, using the vanishing of $F_{\parallel}^{+\mu}$ and a generalized Coulomb condition~\citep{Treat:1972smc}, $A_{\perp}^{\mu}$ has been solved explicitly. It reduces to $A^{\mu}$ in the Coulomb gauge, and approaches $A_{\text{phys}}^{\mu}$ in the IMF limit~\citep{Ji:2013fga,Hatta:2013gta}:
\beq\label{eq:IMFlimit}
A_{\perp}^{\mu,a}(\xi)\approx A^{\mu,a}(\xi)-\frac{1}{\partial^{+}}[(\partial^{\mu}A^{+,b})\mathcal{L}^{ba}(\xi^{\prime-},\xi^{-})]= A^{\mu,a}_{\text{phys}}(\xi).
\eeq
Note that not all the proposed $A_{\perp}^{\mu}$ approach $A_{\text{phys}}^{\mu}$ in the large momentum frame, a counter example is the $A_{\perp}^{\mu}$ constructed from the Lorentz covariant constraints in Refs.~\citep{Delbourgo:1986wz,Guo:2012wv,Guo:2013jia}.

To show the relation between $\big<K^{\mu}\big>$ in the Coulomb gauge and in the lightcone gauge, we adopt the following strategy. First we replace the gluon field in $K^{\mu}$ with $A_{\text{phys}}^{\mu}$ and $A_{\perp}^{\mu}$ (r.h.s and l.h.s of Eq.~\eqref{eq:IMFlimit}) separately, leading to new operators denoted as $K_{A_{\text{phys}}}^{\mu}$ and $K_{A_{\perp}}^{\mu}$. As can be seen from Eq.~(\ref{eq:IMFlimit}), they differ by a power-suppressed contribution under a large Lorentz boost. At this stage, $K_{A_{\perp}}^{\mu}$ and $K_{A_{\text{phys}}}^{\mu}$ are nonlocal operators, and are difficult to realize on the lattice. However, they reduce to local operators 
in Coulomb gauge and lightcone gauge, respectively. Thus, we conclude that for external states with large momentum, $\big<K^{\mu}\big>$ in these two gauges agree at leading power, as shown below:
\begin{equation} \label{eq:method}
    \begin{array}{ccc}
K_{A_{\perp}}^{\mu} & \stackrel{\nabla\cdot A=0}{\longrightarrow} & K^{\mu}\big|_{\nabla\cdot A=0}\\
\quad\downarrow\text{IMF} &  & \quad\downarrow\text{IMF}\\
K_{A_{\text{phys}}}^{\mu} & \stackrel{A^{+}=0}{\longrightarrow} & K^{\mu}\big|_{A^{+}=0}
\end{array}
\end{equation}
\begin{equation} \label{eq:compare}
\big<P'S\big|K^{\mu}\big|PS\big>_{\nabla\cdot A=0}=\big<P'S\big|K^{\mu}\big|PS\big>_{A^+=0}+h.t..
\end{equation}
Eq.~\eqref{eq:compare} paves the way for calculating the total gluon helicity in lattice implementable Coulomb gauge.

In the Coulomb gauge, the gluon field is no longer covariant under Lorentz transformation. The most general decomposition of $\big<P'S\big|K^{\mu}\big|PS\big>$ can be written as:
\begin{align} \label{eq:Coulomb_pri}
\big<P'S\big|K^{\mu}\big|PS\big>_{\nabla\cdot A=0}=&\big<P'S\big|K^{\mu}\big|PS\big>_{\nabla\cdot A=0,\text{finite}}+\big<P'S\big|K^{\mu}\big|PS\big>_{\nabla\cdot A=0,\text{pole}},  \\
\big<P'S\big|K^{\mu}\big|PS\big>_{\nabla\cdot A=0,\text{finite}}=&S^{\mu}a_{1}+\overrightarrow{S}^{\mu}a_{2}+\overrightarrow{\mathbf{P}}^{\mu}a_{3}+\overrightarrow{q}^{\mu}a_{4},  \\ 
 \big<P'S|K^{\mu}\big|PS\big>_{\nabla\cdot A=0,\text{pole}}=&\frac{S^{\mu}}{S\cdot q}b_{1}+\frac{\overrightarrow{S}^{\mu}}{\overrightarrow{S}\cdot\overrightarrow{q}}b_{2}+\frac{\overrightarrow{\mathbf{P}}^{\mu}}{\overrightarrow{\mathbf{P}}\cdot\overrightarrow{q}}b_{3}+\frac{\overrightarrow{q}^{\mu}}{\overrightarrow{q}^{2}}b_{4}, 
\end{align}
where $\mathbf{P}^{\mu}=(P^{\mu}+P'^{\mu})/2$. $\big<P'S\big|K^{\mu}\big|PS\big>_{\nabla\cdot A=0,\text{finite}}$ in Eq.~\eqref{eq:Coulomb_pri} is finite and has a unique value in the forward limit, while $\big<P'S\big|K^{\mu}\big|PS\big>_{\nabla\cdot A=0,\text{pole}}$ contain massless pole(s) as mentioned below Eq.~\eqref{eq:general}. $a_i,b_i$ not only depend on scalar products, but also depend on $\overrightarrow{\mathbf{P}},\overrightarrow{q},\overrightarrow{S}$, generally.
 For $V^{\mu}=S^{\mu},\mathbf{P}^{\mu}$ and $q^{\mu}$, once $V^2$ is fixed, one may choose any four independent elements from the sets $\big\{\overrightarrow{V}^{\mu} \big\}$ and $\big\{V^{\mu}\big\}$ as independent Lorentz structures in the decomposition of $\big<P'S\big|K^{\mu}\big|PS\big>_{\nabla\cdot A=0}$. Different choices are equivalent. Without loss of generality, we choose $\big\{S^{\mu},\overrightarrow{S}^{\mu},\overrightarrow{\mathbf{P}}^{\mu},\overrightarrow{q}^{\mu}\big\}$ for the following discussion, the only ingredient needed is the boost property of each term in Eq.~\eqref{eq:Coulomb_pri}.

In the lightcone gauge, we have ~\cite{Balitsky:1991te}:
\begin{equation} ~\label{eq:axial}
    \big<P'S\big|K^{\mu}\big|PS\big>_{A^+=0}=4S^{\mu}\Delta G(n,q,\mathbf{P})+\frac{n^{\mu}}{n\cdot q}(-4(q\cdot S)\Delta G(n,q,\mathbf{P})+\frac{i}{2}\big<P'S\big|F^{\mu\nu,a}\tilde{F}_{\mu\nu}^{a}\big|PS\big>\big),
\end{equation}
with $\int_0^{\infty} d\xi^-\big<PS\big|F^{n\nu}(\xi^-)F_{\nu}^{\mu}(0)\big|PS\big>=2S^{\mu}\Delta G(n,q,\mathbf{P})$, the gauge-link and color indices are omitted for brevity. $\Delta G(n,0,\mathbf{P})$ is the total gluon helicity $\Delta G$.

When the momentum is large, the $0_{\text{th}}$ and $z_{\text{th}}$ component of $\big<P'S\big|K^{\mu}\big|PS\big>_{\nabla\cdot A=0}$ differ by a power-suppressed quantity. Moreover, for the $0_{\text{th}}$ component,  
the terms proportional to $\vec{V}^{\mu}$ vanish. We thus conclude that:
\begin{align}
\label{eq:Coulomb_fin}
\big<P'S|K^{\mu}\big|PS\big>_{\nabla\cdot A=0,\text{finite}}=&S^{\mu}a_{1}+\text{h.t.}, \nonumber \\
\big<P'S|K^{\mu}\big|PS\big>_{\nabla\cdot A=0,\text{pole}}=& \frac{S^\mu}{S\cdot q}b_1+\text{h.t.} \nonumber \\
=&\frac{n^{\mu}}{n\cdot q}(q\cdot S)(-4\Delta G(n,q,\mathbf{P})+\frac{i}{2}\kappa(q^2))+\text{h.t.},
\end{align}
from the second line to the third line, we have used Eq.~\eqref{eq:compare} and $\big<P'S\big|F^{\mu\nu}\tilde{F}_{\mu\nu}\big|PS\big>=(q\cdot S)\kappa(q^2)$~\citep{Jaffe:1989jz}.

Actually, when the longitudinal components of $P$ and $P'$ are large, the pole part of Eq.~\eqref{eq:Coulomb_pri} can be safely neglected, which can be easily seen from the last line in Eq.~\eqref{eq:Coulomb_fin}. $q\cdot S$ is a Lorentz scalar, while $\kappa(q^2)$, $\Delta G(n,q,\mathbf{P})$ are dimensionless quantities: $\kappa(q^2)$ only depends on Lorentz scalars, $\Delta G(n,q,\mathbf{P})$ is of $\mathcal{O}(1)$ in the power expansion (see Eq.~\eqref{eq:Aphys}). Thus under a boost along $z$ direction, the last line in Eq.~\eqref{eq:Coulomb_fin} gets suppressed due to the increasing $q^+$ in the denominator. Furthermore, the pole part has no impact on the forward limit of Eq.~\eqref{eq:Coulomb_pri}, and thus the latter is finite if we take $q^{\mu}\rightarrow 0$ along the direction $q^+\gg \{q_\perp, q^-\}$. 

From the above discussions, for large $P_z$ and $P'_z$, the Coulomb gauge matrix element reads:
\begin{equation} \label{Coulomb_final}
    O^{\mu}\equiv\big<P'S\big|K^{\mu}\big|PS\big>_{\nabla\cdot A=0}=4S^{\mu}\Delta G(n,q,\mathbf{P})+\text{h.t.},
\end{equation}
where $\mu$ can take either ``$0_{\text{th}}$" or ``$z_{\text{th}}$" component. The lattice calculation of $O^{\mu}$ is feasible, with large momentum external states. In the practical implementation, the matrix element $\big<P'S\big|K^{\mu}\big|PS\big>_{\nabla\cdot A=0}$ with different $q^{\mu}$(down to $q^2\sim$ several hundred $\text{MeV}^2$) can be calculated. An extrapolation of data points to the forward limit will give us the desired total gluon helicity, with controllable power corrections. 

Note that the form of the pole term is intrinsically related to the operator structure of $K^{\mu}$ and the specific gauge used, rather than the kinematics of the external states. In the lightcone gauge, by rewriting the three-gluon term as integrals of field strength tensors, the separated integral after momentum shift will give a $1/(n\cdot q)$ pole~\citep{Balitsky:1991te}. In the non-covariant gauge and some specific kinematics, the forward limit of the pole term has a finite value (negligible when the momentum is approximately lightlike), while the direct forward matrix element of $K^{\mu}$ is ill-defined, independent of which frame we choose.

The bare result of $O^{\mu}$ contains UV divergence that needs to be subtracted, here we use the standard RI/MOM renormalization for local operators. Since the global configuration of the gluon field is of nonperturbative origin, for the on-shell external state, $\big<PS\big|K^{\mu}\big|PS\big>$ is gauge invariant perturbatively. We have checked this explicitly with massive quark state and massless gluon state: 
$$\big<q(P),S\big|K^{\mu}\big|q(P),S\big>,\ \big<g(P),S\big|K^{\mu}\big|g(P),S\big>$$ are independent of the gauge $i$ at one-loop, with $i\in \big\{ A^+=0,A^z=0,\text{general covariant gauge}\big\}.$ Therefore, for the calculation of the renormalization factor, we can choose the off-shell forward matrix element $O^{\mu}_{\text{RI/MOM}}$ in the Landau gauge:
\begin{equation}
O^{\mu}_{\text{RI/MOM}}=\big<i(P),S|K^{0/z}\big|i(P),S\big>_{\partial\cdot A=0,P^{2}=-\mu_{R}^{2},P^{z}=P_{R}^{z}},\ \ \text{i=q,g},
\end{equation}
which captures the same UV-divergent part as $O^{\mu}$.

\subsection{Renormalization in the RI/MOM scheme}
In this section, we give details for the renormalization of the bare, non-forward $\big<K^{\mu}\big>$, in the Coulomb gauge. 

Due to quantum effects, the topological current will mix with other local operators. Under a gauge transformation, $K^{\mu}$ changes by a total derivative Eq.~\eqref{eq:gauge_K}. Thus in general, it is allowed to mix with total derivative operators of the form $\partial^{\mu}O$. $\partial^{\mu}O$ vanishes when sandwiched between forward states, which is the case in our RI/MOM renormalization.
Thus from Lorentz symmetry and requirement of parity, the only allowed operator to mix with $K^{\mu}$ is $j_5^{\mu}=\bar{\psi}\gamma^{\mu}\gamma_5\psi$. No BRST-exact and equation-of-motion (EOM) operator appears in the renormalization procedure~\citep{Collins:1984xc,Joglekar:1975nu,Joglekar:1976eb,Joglekar:1976pe}.

The forward matrix element of $j_{5}^{\mu}$ gives the quark helicity:
\begin{equation} \label{eq:quarkhe}
\big<PS\big|j^{0/z}_5\big|PS\big>_{A^+=0}=S^{0/z}\Delta\Sigma.
\end{equation}
Since the quark helicity is gauge-invariant, so one is free to calculate the bare quark helicity either in the Landau gauge or the Coulomb gauge. The RI/MOM renormalization will be performed in the Landau gauge.
 
As discussed in the previous section, the forward limit of $\big<K^{\mu}\big>$ is the total gluon helicity (after subtracting the higher-twist contaminations): $S^{0/z}\Delta G= \text{lim}_{q^{\mu}\rightarrow0}\big<P-q,S|K^{0/z}\big|PS\big>_{B,\nabla\cdot A=0}$. So what we actually consider is the mixing between gluon helicity and quark helicity in the RI/MOM scheme, we have:
\begin{equation}
\label{eq:RI/MOM}
    \Bigg(\begin{array}{c}
\Delta G_{\text{R}}^{\text{RI}}\\
\Delta \Sigma_{\text{R}}^{\text{RI}}
\end{array}\Bigg)=\Bigg(\begin{array}{cc}
Z_{11}^{\text{RI}} & Z_{12}^{\text{RI}}\\
Z_{21}^{\text{RI}} & Z_{22}^{\text{RI}}
\end{array}\Bigg)\Bigg(\begin{array}{c}
\Delta G_B\\\Delta\Sigma_B
\end{array}\Bigg),
\end{equation}
where $\Delta G_B$ ($\Delta\Sigma_B$) is the bare gluon helicity (quark helicity) calculated on lattice in the Coulomb gauge.
$Z_{ij}^{\text{RI}}$s are defined to subtract all contributions in the off-shell partonic matrix element beyond tree-level, at a specific renormalization point ${P^2=-\mu^2_R,P^z=P_R^z}$:
\begin{equation} \label{def_RI/MOM}
    \Bigg(\begin{array}{c}
\big<PS|K^{0/z}\big|PS\big>_{\text{tree}}\\
\big<PS\big|J_{5}^{0/z}\big|PS\big>_{\text{tree}}
\end{array}\Bigg)=\Bigg(\begin{array}{cc}
Z_{11}^{\text{lattice/}\text{DR}} & Z_{12}^{\text{lattice/}\text{DR}}\\
Z_{21}^{\text{lattice/}\text{DR}} & Z_{22}^{\text{lattice/}\text{DR}}
\end{array}\Bigg)\Bigg(\begin{array}{c}
\big<PS|K^{0/z}\big|PS\big>_{\text{covariant},P^2=-\mu^2_R,P^z=P^z_R}^{\text{lattice/}\text{DR}}\\
\big<PS\big|J_{5}^{0/z}\big|PS\big>_{\text{covariant},P^2=-\mu^2_R,P^z=P^z_R}^{\text{lattice/}\text{DR}}
\end{array}\Bigg),
\end{equation}
where one can either use dimensional regularization or lattice regulator,
the renormalized result (l.h.s of Eq.~\eqref{eq:RI/MOM}) will be independent of the regulator used, but in general has gauge dependence due to the off-shellness of the external state Eq.~\eqref{def_RI/MOM}. Since the total gluon helicity is defined in the $\overline{\text{MS}}$ scheme, after performing RI/MOM renormalization on lattice, one needs the following relation to convert the results to $\overline{\text{MS}}$ scheme: 
\begin{equation}
    \Bigg(\begin{array}{c}
S^{0/z}\Delta G_{\text{R}}^{\text{\ensuremath{\overline{\text{MS}}}}}\\
S^{0/z}\Delta \Sigma_{\text{R}}^{\text{\ensuremath{\overline{\text{MS}}}}}
\end{array}\Bigg)=\Bigg(\begin{array}{cc}
Z_{11}^{\overline{\text{MS}}} & Z_{12}^{\overline{\text{MS}}}\\
Z_{21}^{\overline{\text{MS}}} & Z_{22}^{\overline{\text{MS}}}
\end{array}\Bigg)\Bigg(\begin{array}{cc}
Z_{11}^{\text{RI,DR}} & Z_{12}^{\text{RI,DR}}\\
Z_{21}^{\text{RI,DR}} & Z_{22}^{\text{RI,DR}}
\end{array}\Bigg)^{-1}\Bigg(\begin{array}{c}
S^{0/z}\Delta G_{\text{R}}^{\text{RI,lattice}}\\
S^{0/z}\Delta \Sigma_{\text{R}}^{\text{RI,lattice}}
\end{array}\Bigg),
\end{equation}
The conversion factor $R^{\overline{\text{MS}},\text{RI}}$ is defined as:
$$\Bigg(\begin{array}{cc}
R_{11}^{\overline{\text{MS}},\text{RI}} & R_{12}^{\overline{\text{MS}},\text{RI}}\\
R_{21}^{\overline{\text{MS}},\text{RI}} & R_{22}^{\overline{\text{MS}},\text{RI}}
\end{array}\Bigg)=\Bigg(\begin{array}{cc}
Z_{11}^{\overline{\text{MS}}} & Z_{12}^{\overline{\text{MS}}}\\
Z_{21}^{\overline{\text{MS}}} & Z_{22}^{\overline{\text{MS}}}
\end{array}\Bigg)\Bigg(\begin{array}{cc}
Z_{11}^{\text{RI,DR}} & Z_{12}^{\text{RI,DR}}\\
Z_{21}^{\text{RI,DR}} & Z_{22}^{\text{RI,DR}}
\end{array}\Bigg)^{-1},$$
 $R_{ij}^{\overline{\text{MS}},\text{RI}}$ can eliminate the gauge dependence in $Z_{ij}^{\text{RI,DR}}$ order-by-order. 

$Z_{ij}^{\text{RI,DR}}$s are determined by the following renormalization conditions:
$$Z_{11}^{\text{RI,DR}}=-\frac{\big<g(P),S\big|K^{0/z}\big|g(P),S\big>^{(1)}_{P^2=-\mu^2_R,P^z=P^z_R}}{\big<g(P),S\big|K^{0/z}\big|g(P),S\big>^{(0)}},$$
$$Z_{12}^{\text{RI,DR}}=-\frac{\big<q(P),S\big|K^{0/z}\big|q(P),S\big>^{(1)}_{P^2=-\mu^2_R,P^z=P^z_R}}{\big<q(P),S\big|J_{5}^{0/z}\big|q(P),S\big>^{(0)}},$$
$$Z_{21}^{\text{RI,DR}}=-\frac{\big<g(P),S\big|J_5^{0/z}\big|g(P),S\big>^{(1)}_{P^2=-\mu^2_R,P^z=P^z_R}}{\big<g(P),S\big|K^{0/z}\big|g(P),S\big>^{(0)}},$$
\begin{equation}
Z_{22}^{\text{RI,DR}}=-\frac{\big<q(P),S\big|J_5^{0/z}\big|q(P),S\big>^{(1)}_{P^2=-\mu^2_R,P^z=P^z_R}}{\big<q(P),S\big|J_{5}^{0/z}\big|q(P),S\big>^{(0)}}.
\end{equation}

The calculation of $R_{ij}^{\overline{\text{MS}},\text{RI}}$ is straightforward, we directly give results in the Landau gauge at $\mathcal{O}(\alpha_s)$:
\begin{align} \label{eq:rimom_result}
R_{11}^{\overline{\text{MS}},\text{RI}}(\mu_{R}^{2},\mu^{2})&=1+a_s\Big[\beta_{0}\ln\Big(\frac{\mu_{R}^{2}}{\mu^{2}}\Big)-\frac{367}{36}C_{A}+                          \frac{10}{9}n_{f}\Big],\nonumber \\
R_{12}^{\overline{\text{MS}},\text{RI}}(\mu_{R}^{2},\mu^{2})&=a_s C_F\Big[3\ln\Big(\frac{\mu_{R}^{2}}{\mu^{2}}\Big)-6\Big], \nonumber \\
R_{21}^{\overline{\text{MS}},\text{RI}}(\mu_{R}^{2},\mu^{2})&=-2a_s C_F, \nonumber \\
R_{22}^{\overline{\text{MS}},\text{RI}}(\mu_{R}^{2},\mu^{2})&=1,
\end{align}
where $\beta_0=\frac{11}{3}C_A-\frac{2}{3}n_f$, is the coefficient of the QCD beta function at $\mathcal{O}(a_s)$. $R_{1i}^{\overline{\text{MS}},\text{RI}} (i=1,2)$ contains logarithms of the ratio of RI/MOM scale and $\overline{\text{MS}}$ scale, the coefficients of the logarithms are consistent with the DGLAP evolution of the total gluon helicity~\citep{Altarelli:1977zs,Ji:1995cu}. Interestingly, though free from the renormalization scale dependence, $R_{21}^{\overline{\text{MS}},\text{RI}}$ still contains a finite contribution $-2a_s C_F$, since the on-shell limit and the loop integral are not interchangeable. No new Lorentz structure appears in the off-shell calculation of $K^{\mu}$, which justifies our mixing pattern Eq.~\eqref{eq:RI/MOM}.
Once the renormalized results are converted to the $\overline{\text{MS}}$ scheme, no matching procedure is needed to extract the total gluon helicity.

\section{Gluon helicity from nonlocal gluon correlators}
\label{SEC:matchnonforward}
The strategy of using local operators to calculate the total gluon helicity involves gauge-dependent matrix elements. 
In this section, we propose another method starting from the gauge-invariant gluon helicity quasi-LF correlators, and show how the total gluon helicity can be obtained without a perturbative matching.  

\subsection{Quasi-LF correlators for the extraction of $\Delta G$ }
To extract the total gluon helicity $\Delta G$, one can start from the polarized gluon distribution and take its first moment. The LF operator defining the gluon helicity distribution is given by
\begin{align}
m^{+\mu;+\mu}&=F^{+\mu}(\xi^-)\mathcal{L}(\xi^-,0)\tilde{F}^{+}_\mu(0).
\end{align}
Hereafter color indices of the gluon fields are omitted for brevity. Due to the antisymmetry of the gluon field strength tensor and the Levi-Civita tensor, the polarized gluon distribution operator must satisfy antisymmetry, too~\cite{Balitsky:2021cwr}. The LF gluon operator above is nonlocal and along the lightcone direction, and cannot be directly calculated on the lattice. However, according to LaMET, we can consider a quasi-LF gluon operator which approaches the LF operator in the IMF limit. Here we again take the following quasi-LF operator as an example
 \begin{align}\label{eq:gluon_operator_demo}
&m^{3\mu;0\mu}=F^{3\mu}(z)\mathcal{L}(z,0)\tilde{F}^{0}_\mu(0).
\end{align}
In the literature~\cite{Balitsky:2021cwr,Zhang:2018diq}, there also exist other forms of quasi-LF gluon operators that lead to the same LF physics in the IMF limit. We will briefly discuss them at the end of this section.
 
The quasi-LF correlation defined by $m^{3\mu;0\mu}$ reads
 \begin{align}
\tilde{h}(z,P_z,1/a)=\langle P S| F^{3\mu}(z)\mathcal{L}(z,0)\tilde{F}^{0}_\mu(0) | PS\rangle,
\end{align}
where we have inserted the $1/a$ dependence with $a$ being the lattice spacing to indicate the potential UV divergence on the lattice. $| PS\rangle$ denotes a longitudinally polarized proton state with momentum $P$ and spin $S$. According to the definitions in Eqs.~\eqref{eq:deltag} and~\eqref{eq:deltaG}, we have
\begin{align}
\Delta\tilde{g}(x,P_z,1/a)&=\frac{i}{2xP_z}\int\frac{dz}{2\pi}e^{ix\,zP_z}\,\tilde{h}(z,P_z,1/a),\\
\Delta \tilde{G}(P_z,1/a)&\equiv\int dx\, \Delta \tilde{g}(x,P_z,1/a)=\frac{1}{2P_z}\int_0^\infty{dz}\,\tilde{h}(z,P_z,1/a),\label{eq:gluon_helicity_h}
\end{align}
where we have used
\begin{align}
\frac{1}{x}e^{i x \lambda}=i\int_{\infty(-1+i\epsilon)}^\lambda d\zeta\, e^{i x \zeta },
\end{align}
with $\lambda=zP_z$ being the quasi-LF distance. From the relation in Eq.~\eqref{eq:gluon_helicity_h}, $\Delta \tilde{G}$ can be expressed as an integral over the quasi-LF correlation. Note that at large momentum, the quasi-LF correlation is expected to decay fast enough with $\lambda$ so that the integral is convergent. A relation similar to Eq.~\eqref{eq:gluon_helicity_h} also exists between the total gluon helicity and the LF correlation. 

\subsection{Implementation of renormalization}
The discussion in the previous subsection has been focused on the bare quantities only. They need to be nonperturbatively renormalized in a proper scheme. 
A convenient choice in the literature is the hybrid scheme~\cite{Ji:2020brr}, where one renormalizes the quasi-LF correlation at short distances by a ratio scheme~\cite{Radyushkin:2018cvn}, while uses a self-renormalization scheme~\cite{LatticePartonCollaborationLPC:2021xdx} at long distances.
In this way, the renormalized gluon quasi-LF correlation reads
\begin{align}\label{eq:hybridscheme}
\tilde{h}^{\rm{hyb.}}_R(z,P_z)
=&\frac{\tilde{h}(z,P_z,1/a)}{\tilde{h}(z,P_z=0,1/a)}\theta(z_s-|z|) +	T_s \text{e}^{-\delta m |z|}\,\tilde{h}(z,P_z,1/a)\,\theta(|z|-z_s), 
\end{align}
where $z_s$ is a truncation point within the perturbative region to divide the long-distance and short-distance regions. 
$T_s$ removes the logarithmic divergences in the quasi-LF correlation and ensures its continuity at the truncation point $z=z_s$. $e^{-\delta m |z|}$ is the Wilson line mass renormalization factor, which contains a linearly divergent counterterm $\delta m$ to remove the linear UV divergence arising from the Wilson line self-energy in lattice regularization.

The renormalized quasi-LF correlation in the hybrid scheme can be converted to the $\overline{\rm MS}$ scheme by multiplying with a scheme conversion factor. The reason that we are interested in the $\overline{\rm MS}$ scheme will become clear below. 
In the $\overline{\rm MS}$ scheme, we have
\begin{align}\label{}
\tilde{h}^{\overline{\rm{MS}}}_R(z,P_z,\mu)
=&Z_T(\mu,z,z_s)\tilde{h}^{\rm{hyb.}}_R(z,P_z), 
\end{align}
where $Z_T(\mu,z,z_s)$ is the conversion factor that can be calculated perturbatively. Note that in the hybrid scheme the self-renormalized quasi-LF correlation at long distances are already in the $\overline{\rm MS}$ scheme. Therefore, we need the conversion factor only for short-distance quasi-LF correlations. It can be calculated using the fact that the ratio-renormalized result is independent of the regularization scheme. Thus, we can read out the conversion factor 
from existing results in DR and $\overline{\rm MS}$, which takes the following form at one loop~\cite{Yao:2022vtp}
\begin{align}\label{eq:zeromomgg}
Z_T\left(\mu,z,z_s\right)&=Z\left(\mu,z\right)\theta(z_s-|z|),\nn\\
	Z\left(\mu,z\right)&={\tilde{H}}\left(z,P_z=0,\mu\right)=1+2a_sC_A \left( \frac73 \ln \frac{\mu^2 z^2}{4\text{e}^{-2\gamma_E}} + \frac73 \right)\, ,
\end{align}
where ${\tilde{H}}$ denotes the perturbative quasi-LF correlation.

The $\overline{\rm MS}$ scheme matching kernel for the gluon helicity distribution has been worked out to the NLO, both in coordinate space and in momentum space in Ref.~\cite{Yao:2022vtp}. 
To show their implication, here we write down the coordinate space matching kernel explicitly. Note that in the $\overline{\rm MS}$ scheme, the LF correlation contains scaleless integrals only, and therefore the matching kernel can be straightforwardly read out from $\tilde H$ and takes the following form~\cite{Yao:2022vtp}
\begin{align}
 \bm{C}_{gg}(\alpha,z,\mu)&=\delta(\alpha)+2a_sC_A \bigg\{ \left(4\alpha\bar{\alpha} + 2\left[\frac{\bar{\alpha}^2}{\alpha}\right]_+ \right)\left({\rm{L_z}}-1\right)+6 \alpha\bar{\alpha} -4\left[\frac{\ln(\alpha)}{\alpha}\right]_+ +\left(-3\,{\rm{L_z}}+2\right)\delta(\alpha) \bigg\},\notag\\
\bm{C}_{gq}(\alpha,z,\mu) &=\frac{-2ia_sC_F}{z}\bigg\{ -2\alpha\left(\rm{L_z}+1\right)-4\bar{\alpha} +\rm{L_z}\,\delta(\alpha) \bigg\}\, ,
\end{align}
where $\bar\alpha=1-\alpha$, and ${\rm{L_z}}=\ln\frac{4 e^{-2\gamma_E}}{-\mu^2 z^2}$ from the $\widebar{\rm MS}$ scheme. According to Eq.~\eqref{eq:gluon_helicity_h}, we have
\begin{align}\label{eq:matching_demo}
\Delta \tilde{G}(P_z)
&=\int d\lambda \int_0^1 \frac{d\alpha}{\bar{\alpha}} \left[ \bm{C}_{gg}\left(\alpha,\frac{\lambda}{\bar{\alpha}P_z},\mu\right)h_g(\lambda,\mu)+ \bm{C}_{gq}\left(\alpha,\frac{\lambda}{\bar{\alpha}P_z},\mu\right)h_q(\lambda,\mu)\right]+h.t.= \Delta G+h.t.,
\end{align}
where $h_g(\lambda,\mu)$ and $h_q(\lambda,\mu)$ represent the LF correlations defining the gluon and quark helicity distributions, respectively. Eq.~\eqref{eq:matching_demo} implies that the integral of the polarized gluon matrix element is directly related to the total gluon helicity $\Delta G$ up to power-suppressed contributions, but without a perturbative matching. The same can be shown to be true also in momentum space. 

The demonstration above is based on the NLO result of the matching kernel. However, it is easy to show that
\begin{align} \label{eq:local_form}
&\int_0^\infty dz \langle P S| m^{3\mu;0\mu} | P S\rangle= \langle P S| A^iB^i | P S\rangle|_{A^z=0},
\end{align}
where $\{i,j\}=\{1,2\}$ represent the transverse components, $B^i=\epsilon_{ijk}F^{jk}$ is the color magenetic field. The r.h.s. of Eq.~\eqref{eq:local_form} is the matrix element of the topological current $K^{\mu}$ in the axial gauge, namely, $A^i B^i=K^0\big|_{A^z=0}$. It differs from $\Delta G$ by power suppressed contributions only. 
Thus, we can conclude that the relation Eq.~\eqref{eq:matching_demo} between $\Delta \tilde G$ and the total gluon helicity $\Delta G$ without perturbative matching, is valid to all orders. 
 
As mentioned earlier, the choice for the gluon quasi-LF operator is not unique. For example, the following operator can also be used for lattice calculations of the gluon helicity distribution
 \begin{align}\label{eq:gluon_operator}
&m^{0\mu;3\mu}=F^{0\mu}(z)\mathcal{L}(z,0)\tilde{F}^{3}_\mu(0).
\end{align}
It is easy to prove that the above operator satisfy anti-symmetry~\cite{Balitsky:2021cwr}. 
For this, the no-matching statement above is expected to be true as well, which can be justified by the fact that $\Delta \tilde{G}$ defined in terms of the operator in Eq.~(\ref{eq:gluon_operator}) can be related to the following local matrix element in the temporal gauge:
\begin{align} \label{other_operator}
&\int_0^\infty dz \langle P S|m^{0\mu;3\mu}| P S\rangle= \langle P S| \epsilon_{ij}\left( F^{i0}A^j \right)  | P S\rangle\Big{|}_{A^0=0}, 
\end{align}
%
Since the r.h.s. of Eq.~\eqref{other_operator} is just the matrix element of the topological current $K^{0}$ in the temporal gauge,
we conclude that no matching is required also for the operator in Eq.~\eqref{eq:gluon_operator}. All quasi-LF operators given here can be used to efficiently calculate the total gluon helicity $\Delta G$ from lattice.

\section{Conclusion}
\label{SEC:conclusion}
To conclude, we have presented two approaches to extract the total gluon helicity contribution to proton spin from lattice QCD. The first was based on the nonforward matrix element of the topological current in the Coulomb gauge. We explained the subtlety in taking the forward limit as well as how to implement lattice renormalization in the RI/MOM scheme. The renormalized matrix element was then linked to the total gluon helicity up to power-suppressed contributions. Our second approach was based on nonlocal gluon quasi-LF correlators. We explicitly showed that by choosing appropriate operators, the perturbative matching for the gluon helicity distribution, and thus for the total gluon helicity can be made trivial, provided that all relevant quantities are converted to the $\overline{\rm MS}$ scheme. Our results resolved a long-standing inconsistency in lattice calculations of the total gluon helicity from local and nonlocal operator matrix elements, and have the potential to greatly facilitate these calculations.

\acknowledgments

We thank Yi-Bo Yang for helpful discussions on renormalization and gauge fixing on the lattice. We also thank Shuai Zhao, Yoshitaka Hatta and Aneesh V. Manohar for valuable discussions. This work is supported in part by National Natural Science Foundation of China under grants No. 12375080, 12061131006, 11975051.


\vspace{2em}


\bibliographystyle{apsrev}
\bibliography{ref}

\end{document}